\documentstyle[12pt]{article}

\def\be{\begin{equation}}
\def\ee{\end{equation}}
\def\ben{$$}
\def\een{$$}
\def\ba{\begin{array}{c}}

\def\ea{\end{array}}

\begin{document}

\titlepage
\begin{center}
.

\vspace{2cm}

{\Large \bf
P\"{o}schl-Teller paradoxes
 }\end{center}

\vspace{10mm}

\begin{center}
Miloslav Znojil

\vspace{3mm}

odd\v{e}len\'{\i} teoretick\'{e} fyziky,\\
 \'{U}stav jadern\'e
fyziky AV \v{C}R, 250 68 \v{R}e\v{z},
 Czech Republic\footnote{
paradox.tex, \today; e-mail:
znojil@ujf.cas.cz}\\

\end{center}

\vspace{5mm}

\section*{Abstract}

In the well known P\"{o}schl-Teller trigonometric potential well,
a ${\cal PT}$ symmetric regularization $x \to x - i\,\varepsilon$
of the ``impenetrable" end-point barriers is performed. This leads
to the four different solvable generalizations of the model. As a
byproduct, the scheme clarifies certain apparent paradoxes
encountered in the classically forbidden coupling regime.

\vspace{10mm}

PACS 03.65.Bz 03.65.Ca 03.65.Ge 02.60.Lj 11.30.Er

\newpage

\section{Introduction}

P\"{o}schl-Teller \cite{Poeschl}
potential
 \be
V^{(A,B)}(x) = \frac{A(A-1)}{\cos ^2 x}
+ \frac{B(B-1)}{\sin ^2 x}
 \ee
may be visualised as a sequence of asymmetric wells separated by
impenetrable barriers (cf. Figure~1 where $t=2x/\pi$ and we choose
$A=2.5$ and $B=1.25$).
 Each of these wells
[say, the one defined on the interval $x \in (0, \pi/2)$]
admits the fully non-numerical treatment.
Review paper \cite{Khare}
lists its bound-state spectrum
 \be
E_n=(A+B+2n)^2, \ \ \ \ \ n = 0, 1, \ldots
 \label{clasic}
 \ee
(its three lowest levels are also indicated in Figure~1) as well
as the corresponding wave functions which are proportional to
Jacobi polynomials,
 \be
\psi_n(x) = \cos^A x \cdot \sin^B x \cdot P_n^{(B-1/2,A-1/2)}
\left (
\cos 2x
\right ).
\label{ahas}
 \ee
It is concluded that {\it ``the requirement of $A, B > 0$ \ldots
guarantees"} that each wave function $\psi_n$ in eq. (\ref{ahas})
{\it ``is well behaved and hence acceptable as $x \to 0, \pi/2$"}
(\cite{Khare}, p.  295).  Obviously, the quantized system remains
stable even in the classically collapsing regime with $A(A-1) \in
(-1/4,0)$ [a (weak) barrier re-oriented downwards at the right end
($x=\pi/2$)] or $B(B-1) \in (-1/4,0)$ [same at the left end
($x=0$)]. This is the well known paradox \cite{Landau,Fluegge}.

At $A=B$ we discover another one. The simplified potential does
not change its shape and only its strength varies with $B$, viz.,
$ V^{(B,B)}(x)=g(B)/{\sin ^2 2x}$ where $g(B)=4B\,(B-1)$.  The
ground state is formed at the energy $E_0=4\,B^2=E(B)$.
Nevertheless, at the smallest positive $B <1/2$ the potential well
moves down while, at the same time, the ground-state energy grows.
This contradicts the common sense and represents another paradox
(cf. Figure~2).

An explanation of the latter puzzle is still elementary. It is
sufficient to stay near $x = 0$ and imagine that the differential
Schr\"{o}dinger equation possesses the elementary general
solution,
 \be
\psi(x) \sim C_1\,x^B + C_2\,x^{1-B}, \ \ \ \ \ \ x \approx 0.
 \label{genera}
 \ee
In accordance with the textbooks we would witness the collapse of
the system [i.e., its unstoppable fall in the singularity ${\cal
O}(x^{-2})$] when ${\rm Re\,}B = 1-{\rm Re}\,B$ (i.e., at the
point $B = 1/2$). It is necessary to require that $B > 1/2$. This
means that the domain of $B \in (1/2,1)$ has been counted twice in
Figure~2. The corrected $B-$dependence is displayed in Figure~3.
The decrease and growth of the energy $E_0=E(B)$ reflects merely
the underlying downward and upward move of the potential.
Unfortunately, the whole picture is not entirely satisfactory, at
least for the several purely psychological reasons.

\begin{itemize}

\item
The broken shape of the energy curve in Figure 3 is
counterintuitive.

\item
Although our duty of discarding the well-behaved solutions as
unphysical can find many different mathematical explanations, it
also goes against our basic instincts \cite{Barry}.

\item
People often search for its psychologically persuasive
explanations.  One of the best arguments of the latter type
applies, unfortunately, to the mere {\it regular} $s-$wave
potentials \cite{Newton}.

\item
The fairly popular use of the various {\it ad hoc} conditions can
make the appropriate quantization recipe quite enigmatic
\cite{regul}. The subtleties of its implementation become often
forgotten in quantum chemistry or atomic physics where the
strongly singular phenomenological models frequently occur
\cite{Simsek,Dutra}.

\end{itemize}

\noindent
In what follows we intend to offer a new, unusual approach to
the P\"{o}schl-Tellerian problem, therefore.

\section{Complexification}

In the purely intuitive setting our problem resembles the study of
the elementary algebraic equation $x^2+2b\,x+c=0$ where an
irregularity appears along a parabola $c=c_{crit}(b)= b^2$.
Outside this curve in the real $(b,c)$ plane we always find the
two real roots $x_{1,2}= -b\pm \sqrt{b^2-c}$.  Inside, both of
them suddenly disappear into the complex plane of $x$.

We shall treat the P\"{o}schl-Teller paradoxes in a way guided by
this analogy.

\subsection{${\cal PT}$ symmetric picture}

In essence, the proposal we are going to describe will replace the
previous pictures by a new Figure~4.  Its mathematics will offer
us the two smooth auxiliary curves $E^{(\pm)}(B)$. Their user will
be permitted to make his/her choice between these two
alternatives, employing in addition his/her purely physical
arguments and/or preferences.

The core of our proposal will lie in a certain complexification of
the P\"{o}schl-Teller differential Schr\"{o}dinger equation in
units $2m = \hbar =1$,
 \be
 \left ( -\frac{d^2}{d {x}^2}
+ \frac{\alpha^2-1/4}{\cos^2{x}}
+ \frac{\beta^2-1/4}{\sin^2{x}}
 \right ) {\psi}({x}) = E\,{\psi}({x}), \ \ \ \ \ \ \ \
 {x} \in (0, \pi/2 ),
 \label{angular}
 \ee
with $\alpha = A-1/2 > 0$ and $\beta = B-1/2 > 0$.  We shall be
inspired by the recent papers by Bender et al \cite{BB} who
replace the real (interval of) coordinates $x$ by a suitable
complex curve ${\cal C}(t)$.  On the basis of an extensive
computational and WKB-based experience with many resulting
non-Hermitian Hamiltonians they conjecture that under certain weak
conditions the spectrum can still stay real. In the present
context, the most natural implementation of the latter idea is
based on the elementary choice of the straight line,
 \be
 x \to {\cal C}(t) = x(t) = t - i\,\varepsilon,
\ \ \ \ t \in (-\infty,\infty).
 \label{traj}
 \ee
The trick has been shown to work, e.g., for the shape invariant
models \cite{shape} and for the whole Natanzon class of the
exactly solvable potentials \cite{Geza}.

One can easily check that the curve (\ref{traj}) remains unchanged
under the combined action of the parity ${\cal P}$ and of the
``time-reversal" (in fact, complex-conjugation) operator ${\cal
T}$.  One can speak about the specific, ${\cal PT}$ symmetric
quantum mechanics \cite{BBjmp}. Under certain mathematical
assumptions and at least in a certain domain of couplings (both
specifications are, by far, not yet clear \cite{Pham}) one can
often discover that the spectrum after deformation $x \to {\cal
C}(t)$ remains discrete, bounded from below and {\em
real}~\cite{others}.

\subsection{Wave functions \label{subs} }

In a way guided by the original papers \cite{Poeschl} let us move
to the complex $x$, abbreviate $\sin^2x = y$, denote $\psi[x(y)] =
\varphi(y)$ and re-write our equation (\ref{angular}) using these
new variables,
 \be
 y(1-y)\,\varphi''(y) + \left( \frac{1}{2}-y\right)\,\varphi'(y)
 +
 \frac{1}{4}\left (
E-
 \frac{\beta^2-1/4}{y}
 -\frac{\alpha^2-1/4}{1-y}
 \right )\,\varphi(y)
 = 0.
\label{gausser}
 \ee
The ansatz $E=k^2$ and
 \ben
 \varphi(y) = y^\mu(1-y)^\nu\,f(y)
 \een
transforms our complexified differential equation in its Gauss
hypergeometric equivalent
 \ben
  y(1-y)\,f''(y) +
  \left [
  \left( 2\mu+ \frac{1}{2} \right)
 - (2\mu+2\nu+1 )\,y
  \right ]
  \,f'(y)
 +
\left [ \frac{1}{4}
 k^2-
 (\mu+\nu)^2
 \right ]\,f(y)
 = 0
 \een
provided only that we choose $\mu$ and $\nu$ in accord with the
conditions
 \ben
 4\mu(\mu-1) = \beta^2-1/4, \ \ \ \ \ \ \ \
 4\nu(\nu-1) = \alpha^2-1/4.
 \een
In terms of the two indeterminate signs these quadratic equations
define the two pairs of the eligible exponents,
 \ben
 2\mu = \frac{1}{2} + \sigma\,\beta=\kappa(\sigma), \ \ \ \
 2\nu= \frac{1}{2} + \tau\,\alpha=\lambda(\tau) , \ \ \ \ \
\sigma,\tau=\pm 1.
 \een
The general solution of our equation (\ref{angular}) acquires the
$\tau-$dependent compact form
 \be
 \psi(x) =
 \left \{
 C_1\,
  \chi^{(\sigma,\tau)}\left [y(x)\right ]
 \sin^{\kappa(\sigma)}x
 +
  C_2\, \chi^{(-\sigma,\tau)}\left [y(x)\right ]
 \sin^{\kappa(-\sigma)}x
  \right \}
  \cdot  \cos^{\lambda(\tau)}x .
  \label{suma}
  \ee
Here we have abbreviated
 \be
 \chi^{(\sigma,\tau)}(y)
 =
 \,_2F_1 \left \{
 \frac{1}{2} \left [\kappa(\sigma) +\lambda(\tau)+k\right ],
 \frac{1}{2} \left [\kappa(\sigma) +\lambda(\tau)-k\right ],
 \frac{1}{2}+\kappa(\sigma);y
 \right \}.
  \label{sumatra}
 \ee
In an alternative representation using {\em the same}
constants $C_1$ and $C_2$ we have
 \be
 \chi^{(\sigma,\tau)}\left [y(x)\right ] \cdot
 \cos^{\lambda(\tau)}x=
 \varrho^{(\sigma,\tau)}\left [y(x)\right ] \cdot
 \cos^{\lambda(\tau)}x
 +
 \varrho^{(\sigma,-\tau)}\left [y(x)\right ] \cdot
 \cos^{\lambda(-\tau)}x
  \label{sumabre}
 \ee
where
 \be
 \varrho^{(\sigma,\tau)}(y)
 =
 G^{(\sigma,\tau)}
 \,_2F_1 \left \{
 \frac{1}{2} \left [\kappa(\sigma) +\lambda(\tau)+k\right ],
 \frac{1}{2} \left [\kappa(\sigma) +\lambda(\tau)-k\right ],
 \frac{1}{2}+\lambda(\tau);1-y
 \right \}
  \label{sumaris}
 \ee
with the factor
 \ben
 G^{(\sigma,\tau)}=
 \frac{\Gamma(1+\sigma \,\beta)\,\Gamma(-\tau\,\alpha)}
 {\Gamma\left [\kappa(\sigma) +\lambda(-\tau)+k\right ]\,\Gamma
 \left [\kappa(\sigma) +\lambda(-\tau)-k\right ]}\ \cdot
 \een
One of the immediate consequences of these two alternative
expansions is our explicit knowledge of the related $0<x \ll 1$
left-threshold leading-order approximation
 \be
 x^{-1/2}\psi(x) \sim
 C_1\, x^{\sigma\,\beta}
 +
  C_2\,
 x^{-\sigma\,\beta}
  \label{sumba}
  \ee
and of its $0< z = {\pi}/{2}-x \ll 1$ right-threshold counterpart
 \be
 z^{-1/2}\psi[x(z)] \sim
\tilde{C}^{(+)}\,
  z^{\tau\, \alpha }
 +
 \tilde{C}^{(-)}\,
  z^{-\tau\, \alpha },\ \ \ \ \
  \label{sumbb}
 \tilde{C}^{(\pm)}=
 \left [ C_1\,G^{(\sigma,\pm \tau)}+ C_2\,G^{(-\sigma,\pm
\tau)} \right ].
 \ee
We may summarize that the complete solution of the
P\"{o}schl-Tellerian differential Schr\"{o}dinger equation is
available in closed form even on any generalized, complex domain
${\cal C}$ of coordinates $x$ characterized, presumably, by a
suitable form of its ${\cal PT}$ symmetry, ${\cal C}={\cal
PT}{\cal C}{\cal PT}$.

\section{Spectra}

The structure of the above general wave functions indicates that
the singularities $x=0$ and $z=\pi/2-x=0$ remain most suitable
points where we can impose the boundary conditions. As long as
they do not belong (by our assumption) to our complex curve of
coordinates ${\cal C}(t)$, the choice and specification of these
boundary conditions is not constraint by any (usually, obligatory)
conditions of regularity.

One should still be careful in the classically forbidden domain of
the very small couplings $\alpha>0$ and $\beta>0$. The related
possible difficulties are well known. They represented a good
reason for the Fl\"{u}gge's unnecessarily restrictive ``safe"
postulates $A>1$ and $B>1$ which he uses in his textbook
(\cite{Fluegge}, p. 89).

\subsection{Refined boundary conditions}

Once we wish to discuss the specific ability of quantum mechanics
which can protect its systems from a collapse into (sufficiently
weakly) attractive singularities, we have to eliminate the
superfluous solutions by all means including the brute force
\cite{Landau,Barry,regul}. In the present context, an application
of the latter rule is significantly facilitated by our explicit
knowledge (\ref{sumba}) and (\ref{sumbb}) of the independent
solutions in the leading-order approximation,
 \be
 \psi(x) \sim x^{1/2 \pm \beta}, \ \ \ \ x \ll 1,
 \ \ \ \ \ \ \ \
 \psi(x) \sim (\pi/2-x)^{1/2 \pm \alpha}, \ \ \ \ x \sim \pi/2.
 \label{obi}
 \ee
Obviously, at the smallest couplings, these estimates remain
compatible with the current boundary conditions
$\psi(0)=\psi(\pi/2)=0$ at an (almost) {\em arbitrary} positive
energy $E$. A new paradox is born. Even in the Hermitian case with
$\varepsilon=0$, the necessary physical re-installation of the
regularity must be achieved via the more restrictive boundary
conditions
 \be
 \lim_{x \to 0}\psi(x)/\sqrt{x} = 0, \ \ \ \ \ \ \ \ \beta \in
(0,1/2)
 \label{sample}
 \ee
(plus, {\it mutatis mutandis}, for $x$ near $\pi/2$; cf. refs.
\cite{Dutra} for another solvable illustration of this rule).

In the present regularized ${\cal PT}$ symmetric generalization
both the components in eq. (\ref{obi}) remain equally acceptable.
The situation is similar to the asymmetric but regular Hermitian
models where one sometimes selects between the Dirichlet and
Neumann (or, in general, mixed) boundary conditions. Here, at the
small couplings, any similar generalized requirements must be
refined as well, working with the limits similar to eq.
(\ref{sample}). Numerically, the situation will be badly
ill-conditioned but we can still start from the fixed initial
values of $C_1$ and $C_2$ at $x=0$ and determine (the discrete set
of) the energies $E_n$ from another postulate of another fixed set
of parameters $\tilde{C}^{(+)}$ and $\tilde{C}^{(-)}$ at
$x=\pi/2$.

Similar ``weakly solvable" models which do not require any
termination of the hypergeometric series also do occur in
applications from time to time \cite{Dutradva}. We are not going
to study them here in any detail.

\subsection{Classification of the exactly solvable cases}

After our present regularization, one has to contemplate the whole
infinite domain of $x(t)$ or $t \in (-\infty,\infty)$. We shall
omit here also this direction of considerations which,
generically, leads to the Floquet theory and to the characteristic
band spectra for the ${\cal PT}$ symmetric and periodic systems
\cite{periodic}.

In a narrower domain of applications related, e.g., to the
attempts to generalize \cite{Quesne} or ${\cal PT}$ symmetrize
\cite{Milos} the Calogero's three-body model \cite{Calogero} we
shall solely pay attention to the problems which keep using the
``physical" boundary conditions imposed directly at the poles at
$x = 0$ and $x = \pi/2$.

There exist several good practical reasons (e.g., the well known
slow convergence of the infinite hypergeometric series) for the
exclusive preference of the terminating, polynomial
P\"{o}schl-Teller solutions. In such a setting, our explicit
knowledge of the general solutions facilitates also the complete
classification of the eligible boundary conditions.

In the first step it is important to notice that for the
superposition (\ref{suma}), generically, the two necessary
termination conditions are mutually incompatible. Fortunately,
they differ just in one sign, $\sigma \to -\sigma$. Without any
loss of generality  we may put $C_2=0$ and write down the general
termination condition, therefore,
 \ben
 k=k^{(\sigma,\tau)}_n=\sigma\,\beta+\tau\,\alpha+2n+1,\ \ \ \ \ \
 n = 0, 1, \ldots .
 \een
It reduces the infinite series (\ref{sumatra}) to the elementary
Jacobi polynomial and, {\em simultaneously}, nullifies the
co-factor $G$ of the second subseries (\ref{sumaris}) in the
alternative formula (\ref{sumabre}). Summarizing, we are left with
the unique elementary solution $\psi(x) =
 \psi^{(\sigma,\tau)}_n(x) $ with the
 energies $E_n^{(\sigma,\alpha)}
 =\left [
 k_n^{(\sigma,\tau)}\right]^2$ and wave functions
 \ben
 \psi^{(\sigma,\tau)}_n(x) =
 C_1\,
  \sin^{1/2+\sigma\,\beta}x\, cos^{1/2+\tau\,\alpha)}x
  \,_2F_1 \left (
 -n, n+1+\sigma\,\beta+\tau\,\alpha,1+\sigma\,\beta;\sin^2x
 \right ).
  \label{sumo}
 \een
Only our choice of the signs $\sigma=\pm 1$ and $\tau=\pm 1$
remains variable. In all these four cases there is no freedom left
for our choice of the boundary conditions. By construction our
solutions simply fit the $x \to 0$ rule
 \be
 x^{-1/2}\psi^{(\sigma,\tau)}(x) =
 C_1\, x^{\sigma\,\beta}
 +
  0 \cdot
 x^{-\sigma\,\beta}
  \label{swumba}
  \ee
and its $x \to \pi/2$ parallel
 \be
 z^{-1/2}\psi^{(\sigma,\tau)}(x) =
 C_1\,
   G^{(\sigma, \tau)}(\pi/2-x)^{\tau\, \alpha }
   + 0 \cdot(\pi/2-x)^{-\tau\, \alpha }.
  \label{swumbb}
 \ee
At every main quantum number $n = 0, 1, 2, \ldots $ we have a
choice among the quadruplet of boundary conditions (\ref{swumba})
+  (\ref{swumbb}) giving the respective four different energy
series
 \ben
 E^{(\sigma,\tau)}_n=\alpha^2+\beta^2 +2\sigma\tau\alpha\beta
 +(4n+2)(\sigma\beta+\tau\alpha) + (2n+1)^2
 \label{energe}
 \een
numbered by $\sigma = \pm 1$ and $\tau = \pm 1$. Two of them [cf.
$E_n^{(+,+)}=\left ( \Sigma +2n+1 \right )^2$ and $ E^{ ( - ,-)}
=\left ( \Sigma -2n-1\right )^2$] depend on the sum
$\Sigma=\alpha+\beta$ and, in this sense, resemble strongly the
Hermitian formula (\ref{clasic}). The other two series
$E_n^{(+,-)}=\left ( \Delta +2n+1 \right )^2$ and $ E^{ ( - ,+)}
=\left ( \Delta -2n-1\right )^2$ exhibit a dependence on the mere
difference $\Delta=\beta-\alpha$ and remain, unexpectedly,
coupling-independent for the symmetric wells $V^{(B,B)}(x)$. In
contrast, as already mentioned (cf. Figure~4 above), an interplay
or superposition of the former two series provides one of the
``most natural" explanations of the paradox in Figures~2 or 3.

\section*{Acknowledgement}

Partially suported by GA AS CR grant No. A
 1048004.

\section*{Figure captions}

\subsection*{Figure 1. P\"{o}schl-Teller potential}

\subsection*{Figure 2. Paradox of review \cite{Khare}}

\subsection*{Figure 3. Corrected picture}

\subsection*{Figure 4.
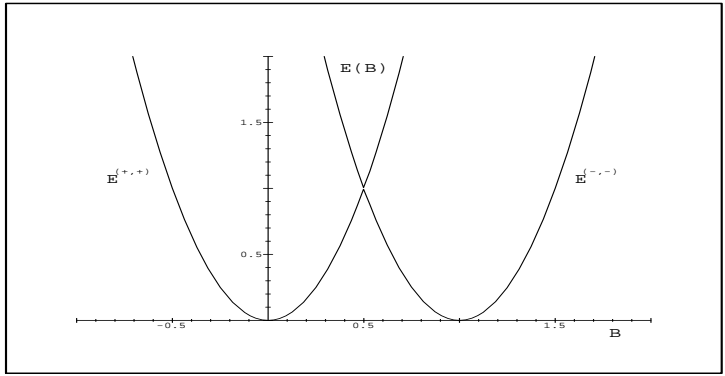
${\cal PT}-$symmetric re-interpretation of Figure 3}


\newpage

 \begin{thebibliography}{00}


\bibitem{Poeschl}
P\"{o}schl G and Teller E 1933 Z. Physik 83 143;

Quesne C 1999 J. Phys. A: Math. Gen. 32 6705 with further
references

\bibitem{Khare}
Cooper F, Khare A and Sukhatme U 1995 Phys. Rep. 251 267

\bibitem{Landau}
Landau L D and Lifshitz E M 1960 Quantum Mechanics (London:
Pergamon Press), ch. V, par. 35.

\bibitem{Fluegge}
Fl\"{u}gge S 1971 Practical quantum mechanics I (Berlin:
Springer), p. 157

\bibitem{Barry}
Reed M and Simon B 1978 Methods of Modern Mathematical Physics IV
(New York: Academic)

\bibitem{Newton}
Newton R G 1982 Scattering Theory of Waves and Particles (Berlin:
Springer), pp. 391-392.

\bibitem{regul}
Frank W M, Land D J and Spector R M 1971 Rev. Mod. Phys. 43 36

\bibitem{Simsek}
Simsek M and Yalcin Z 1994 J. Math. Chem. 16 211;

Znojil M 1996 J. Math. Chem. 19  205;

Chun-Sheng Jia, Jia-Ying Wang, Su He and Linag-Tian Sun 2000 J.
Phys. A: Math. Gen. 33 6993

\bibitem{Dutra}
Znojil M 2000 Phys. Rev. A 61  066101

\bibitem{BB}
Bender C M and Boettcher S 1998 Phys. Rev. Lett. { 24}  5243

\bibitem{shape}
Bagchi B and Roychoudhury R 2000 J. Phys. A: Math. Gen. 33 L1;

Znojil M 2000 J. Phys. {A}: Math. Gen. 33 L61 and 4561

\bibitem{Geza}
L\'{e}vai G and Znojil M 2000 J. Phys. A: Math. Gen. 33 7165

\bibitem{BBjmp}
Caliceti E, Graffi S and Maioli M 1980 Commun. Math. Phys. 75 51;

Buslaev V and Grechi V 1993 J. Phys. A: Math. Gen. 26 5541;

Fern\'andez F M, Guardiola R, Ros J and Znojil M 1998 J. Phys. {
A}: Math. Gen. 31 10105;

Andrianov A A, Ioffe M V, Cannata F and Dedonder J P 1999 Int. J.
Mod. Phys. A 14 2675;

Bender C M, Boettcher S and Meisinger P N 1999 J. Math. Phys. 40
2201;

Mezincescu G A 2000 J. Phys. A: Math. Gen. 33 4911;

Bagchi B and Quesne C 2000 Phys. Lett. A 273  285

\bibitem{Pham}
Delabaere E and Pham F 1998 Phys. Letters A 250 25 and 29;

Bender C M, Berry M, Meisinger P N, Savage V M and Simsek M 2001
J. Phys. { A}: Math. Gen. 34 L31

\bibitem{others}
Cannata F,  Junker G and Trost J 1998 Phys. Lett. { A 246} 219;

Znojil M, Cannata F, Bagchi B and Roychoudhury R 2000 Phys. Lett.
B 483 284

Bender C M, Boettcher S and Savage Van M 2000 J. Math. Phys. 41
6381

Fern\'andez F M, Guardiola R, Ros J and Znojil M 1999 J. Phys. {
A}: Math. Gen. 32 3105;

Bender C M, Boettcher S, Jones H F and Van Savage M 1999 J. Phys.
A: Math. Gen. 32 6771;

Bagchi B, Cannata F and Quesne C 2000 Phys. Lett. A 269 79;

Znojil M and Levai G 2000 Phys. Lett. A 271 327;

L\'{e}vai G, Cannata F and Ventura A 2001 J. Phys. A: Math. Gen.
34 1

\bibitem{Dutradva}
de Souza-Dutra A 2000 Phys. Rev. A 47 (2000) 066102;

Jiu-Xun Sun 1999 Acta Phys. Sinica 48 1992.

\bibitem{periodic}
Bender C M, Dunne G V and  Meisinger P N 1999  Phys. Lett. A 252
272

\bibitem{Quesne}
Khare A and Quesne C 1998 Phys. Lett. A 250 33

\bibitem{Milos}
Znojil M and Tater M 2001 J. Phys. { A}: Math. Gen. 34 1793

\bibitem{Calogero}
Calogero F 1969 J. Math. Phys. 10  2191


\end{thebibliography}
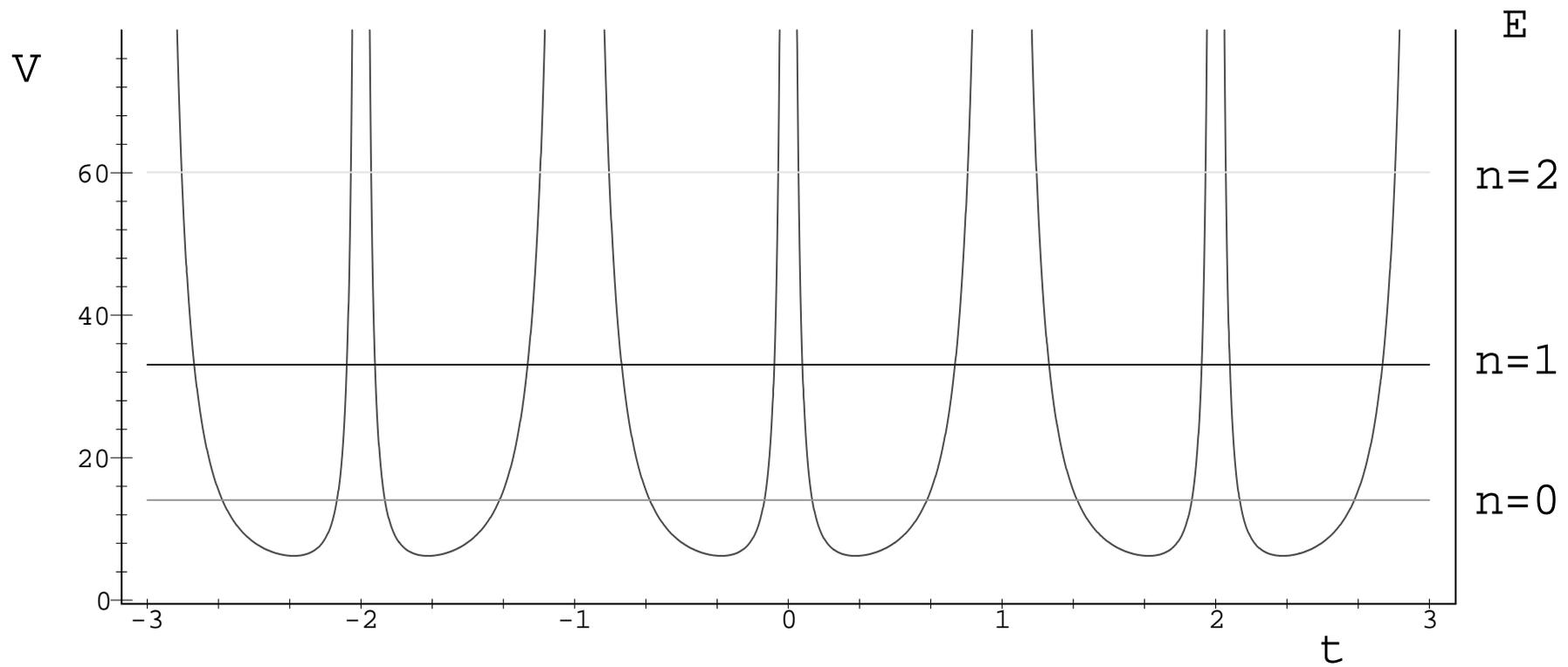
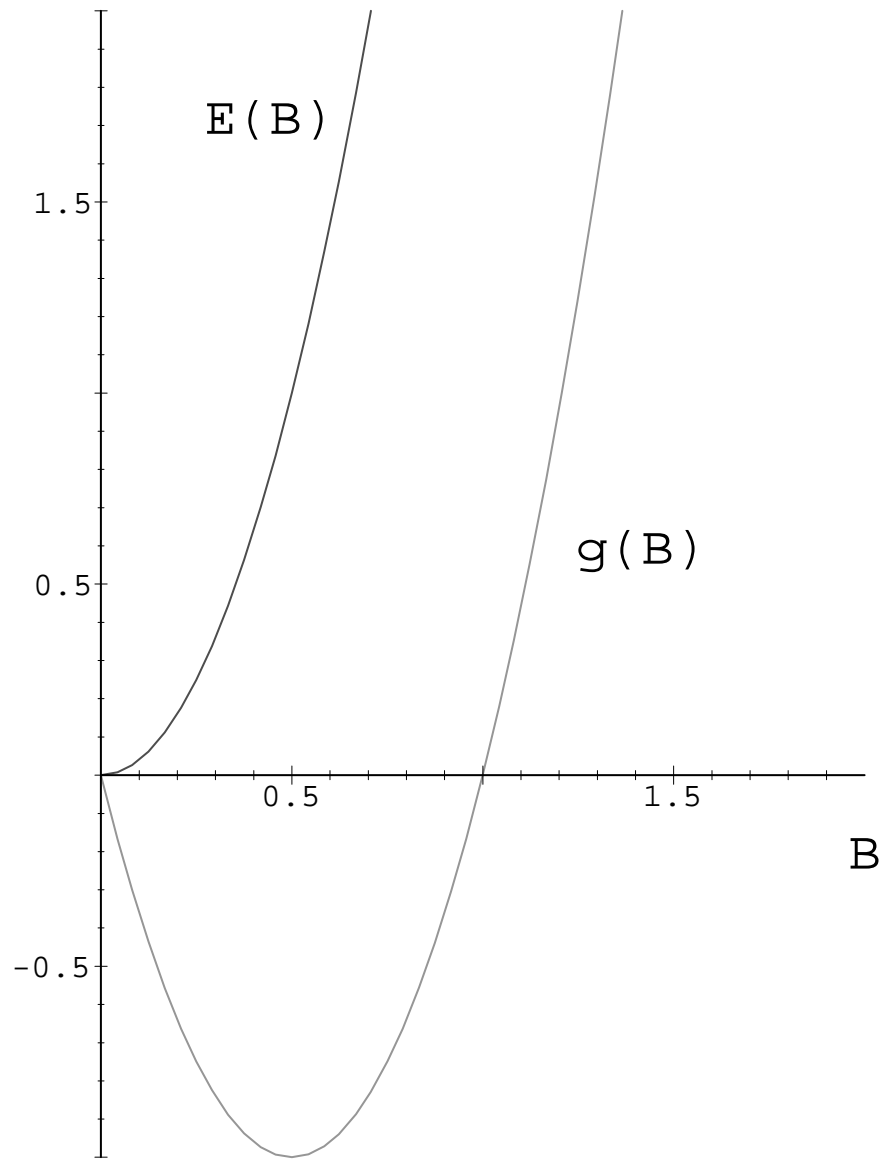
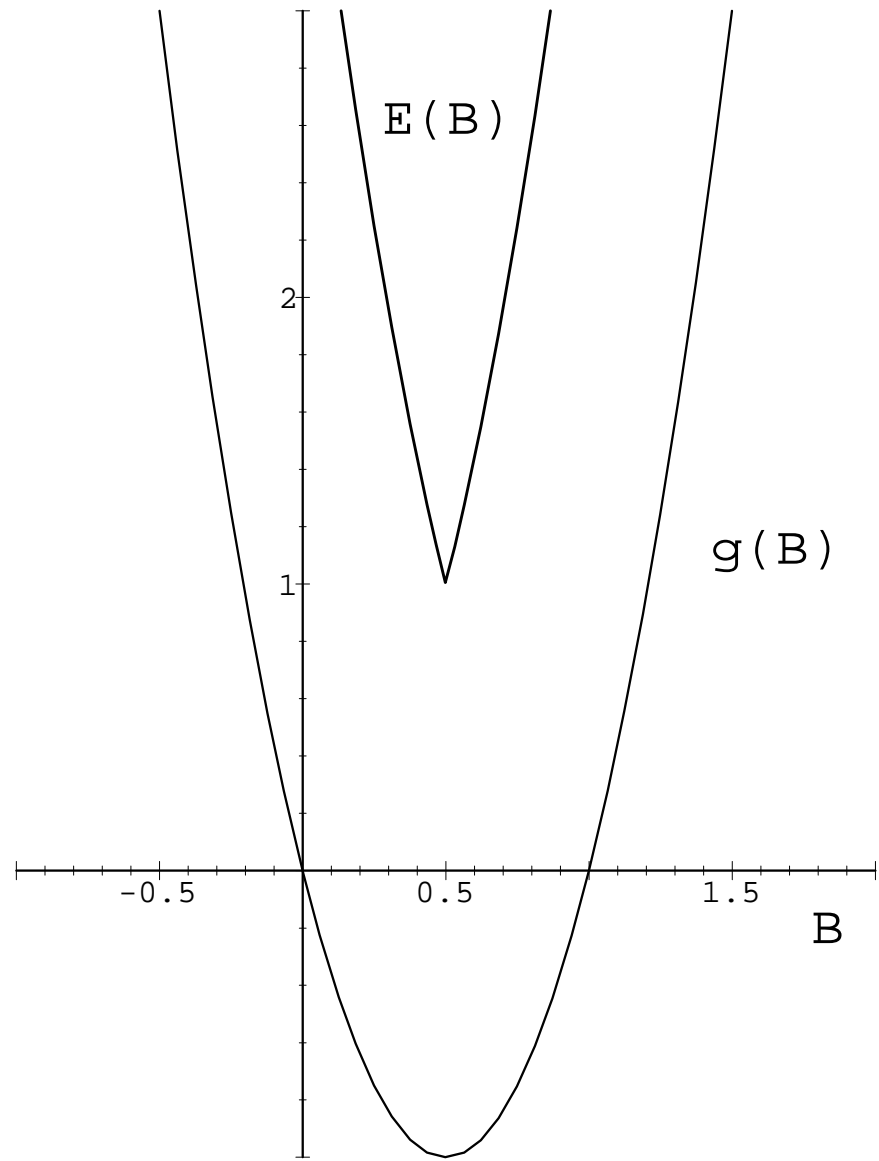
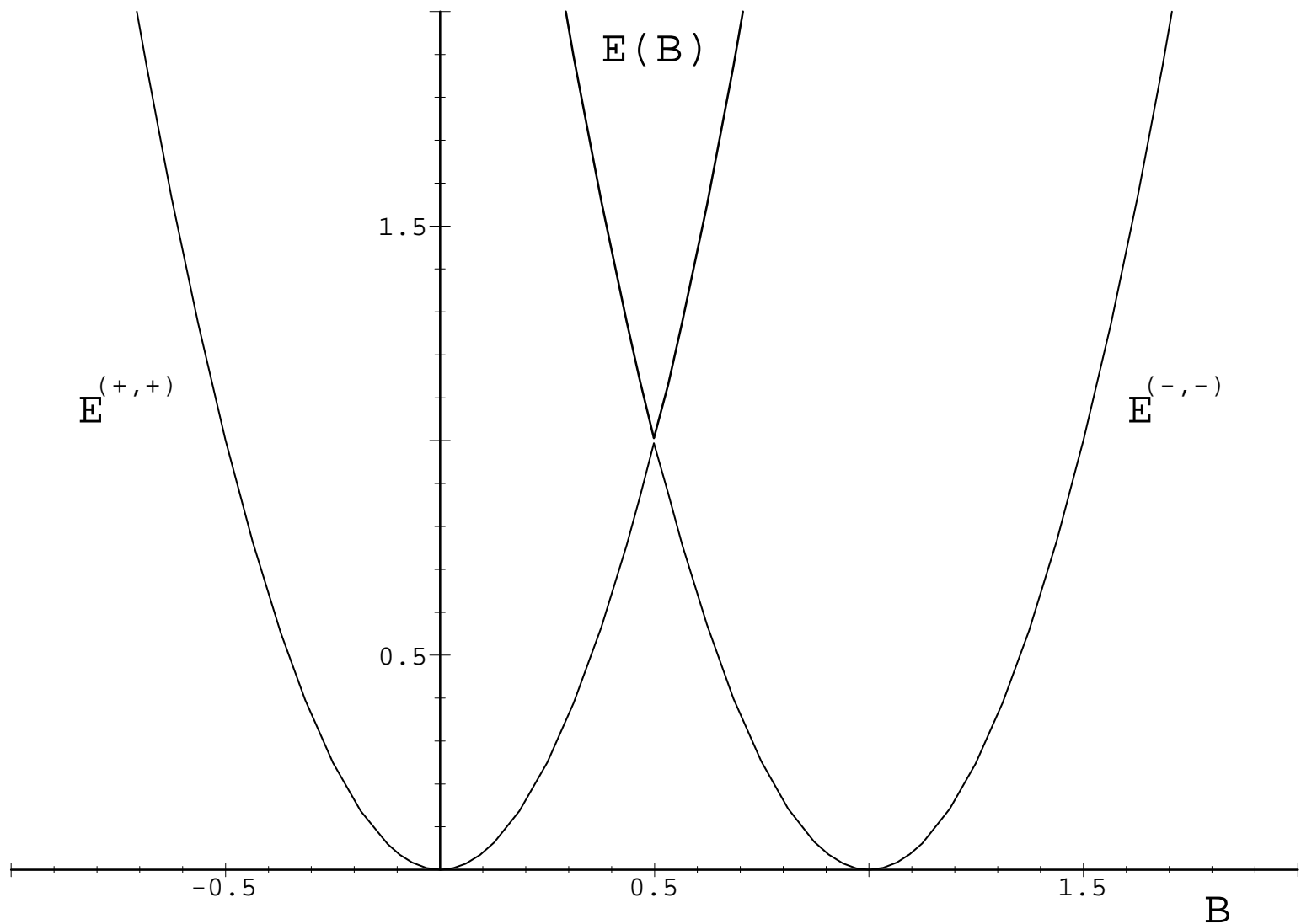
\end{document}